\documentclass[sigconf]{acmart}
\usepackage[T1]{fontenc}
\usepackage{newtxtext}
\usepackage{xspace}
\usepackage{multirow}
\newcommand{\approach}{L\textsc{ive}C\textsc{oder}\xspace}
\usepackage[most]{tcolorbox}
\usepackage{csquotes}
\usepackage[noorphans,vskip=1em,leftmargin=1em]{quoting}
\usepackage{xurl}
\usepackage{fontawesome5}
\usepackage{etoolbox}
\usepackage{algorithm}
\usepackage{algpseudocode}
\usepackage[table]{xcolor}
\usepackage{colortbl}
\definecolor{grey}{gray}{0.95}

\usepackage{amsmath}

\usepackage{booktabs}
\usepackage{multirow}
\usepackage{makecell}
\usepackage{threeparttable} % optional
\usepackage{xcolor}
\usepackage{graphicx} % for \resizebox
\definecolor{ForestGreen}{RGB}{34,139,34}

\usepackage{tabularx}

\usepackage{array}
\usepackage{siunitx}
\usepackage{enumitem}

\setlist[itemize]{
    labelindent=2pt,
    leftmargin=*,
    labelsep=0.5em
}

\definecolor{DeepRed}{RGB}{196,59,59}
\definecolor{LightRed}{RGB}{252,239,239}

\hypersetup{
  colorlinks=true,
  linkcolor=black,
  citecolor=black,
  urlcolor=oai-red,
}

\newcommand{\costabs}[1]{\textcolor{gray}{\scriptsize(#1)}}

\newcommand{\chg}[1]{\textcolor{gray}{\scriptsize\,({#1})}}

\newcommand{\redimp}[1]{\textcolor{DeepRed}{\textbf{#1}}}

\definecolor{navyblue}{HTML}{0071BC}
\definecolor{hotpink}{HTML}{FF0080}

\definecolor{oai-white}{HTML}{FFFFFF}
\definecolor{oai-black}{HTML}{000000}
\definecolor{oai-red}{HTML}{FF4500}
\definecolor{oai-green}{HTML}{51DA4C}
\definecolor{oai-blue}{HTML}{0000FF}
\definecolor{oai-yellow}{HTML}{FFF639}
\definecolor{oai-magenta}{HTML}{FF45FF}
\definecolor{oai-cyan}{HTML}{00FFFF}
\definecolor{oai-orange}{HTML}{FE7600}
\definecolor{oai-violet}{HTML}{8A2BE2}
\definecolor{oai-brown}{HTML}{A0522D}

\definecolor{oai-green-050}{HTML}{F4FFF4}
\definecolor{oai-green-100}{HTML}{E9FFE8}
\definecolor{oai-green-200}{HTML}{D9FFD8}
\definecolor{oai-green-300}{HTML}{C9FFC7}
\definecolor{oai-green-400}{HTML}{A6FFA3}
\definecolor{oai-green-500}{HTML}{7CF178}
\definecolor{oai-green-600}{HTML}{51DA4C}
\definecolor{oai-green-700}{HTML}{3FA93B}
\definecolor{oai-green-800}{HTML}{2D712A}
\definecolor{oai-green-900}{HTML}{193718}

\definecolor{oai-gray-000}{HTML}{FFFFFF}
\definecolor{oai-gray-100}{HTML}{FAFAFA}
\definecolor{oai-gray-200}{HTML}{F5F5F5}
\definecolor{oai-gray-300}{HTML}{E5E5E5}
\definecolor{oai-gray-400}{HTML}{FFB7A4}
\definecolor{oai-gray-500}{HTML}{CDCDCD}
\definecolor{oai-gray-600}{HTML}{A8A8A8}
\definecolor{oai-gray-700}{HTML}{747474}
\definecolor{oai-gray-800}{HTML}{393939}
\definecolor{oai-gray-900}{HTML}{000000}

\usepackage[most]{tcolorbox}
\usepackage{xcolor}
\usepackage{array}

\definecolor{kbBg}{RGB}{245,245,245}
\definecolor{kbBar}{RGB}{180,180,180}
\definecolor{kbNum}{RGB}{150,150,150}
\definecolor{kbText}{RGB}{55,55,55}
\definecolor{kbGreen}{RGB}{95,130,90}
\definecolor{kbBlue}{RGB}{35,85,185}

\newtcolorbox{knowledgebox}{
  enhanced,
  colback=kbBg,
  colframe=kbBg,
  boxrule=0pt,
  sharp corners,
  left=6pt,
  right=6pt,
  top=6pt,
  bottom=6pt,
  borderline west={1.2pt}{0pt}{kbBar},
  before skip=0pt,
  after skip=0pt
}

\definecolor{AlgoComment}{RGB}{170,0,85}   % 紫红色
\definecolor{AlgoPhase}{RGB}{170,0,85}     % 更亮一点的阶段注释色
\definecolor{AlgoGray}{RGB}{90,90,90}

\definecolor{myitem}{HTML}{A64D79}

\newcommand{\cnum}[2]{%
  \tikz[baseline=(C.base)]%
    \node[draw=none, circle, fill=#1, inner sep=1.3pt, minimum size=1em] (C)
    {\color{white}\bfseries\scriptsize #2};%
}

\algrenewcommand\algorithmicrequire{\textbf{Input:}}
\algrenewcommand\algorithmicensure{\textbf{Output:}}

\algrenewcommand\alglinenumber[1]{\footnotesize\bfseries #1}

\algrenewcommand\algorithmiccomment[1]{\hfill{\footnotesize\textcolor{AlgoComment}{\#~#1}}}

\AtBeginDocument{%
  }

 \setcopyright{none}

\newcommand{\inst}[1]{\textsuperscript{#1}}

\author{
Ruwei Pan\inst{1,2} \quad
Jiangshuai Wang\inst{1} \quad
Qisheng Zhang\inst{1} \quad
Yueheng Zhu\inst{1,2} \quad
Linhao Wu\inst{2} \\[2pt]
Zixiong Yang\inst{2} \quad
Yakun Zhang\inst{3} \quad
Lu Zhang\inst{2} \quad
Hongyu Zhang\inst{1}
}

\authornote{
\inst{1} Chongqing University \quad
\inst{2} Peking University \quad
\inst{3} Harbin Institute of Technology (Shenzhen)
}

\begin{document}

% \title{\approach: Cross-Attempt Knowledge Optimization for Repository-Level Code Generation}

\title{Persistent Cross-Attempt State Optimization for Repository-Level Code Generation}

% \affiliation{%

% \affiliation{%

% \affiliation{%

\begin{abstract}

Large language models (LLMs) have achieved substantial progress in repository-level code generation. However, solving the same repository-level task often requires multiple attempts, while existing methods still optimize each attempt in isolation and do not preserve or reuse task-specific state across attempts.
In this paper, we propose \approach, a novel framework for repository-level code generation based on cross-attempt knowledge optimization. \approach maintains persistent task-specific state from prior attempts to guide subsequent generation. This state includes success knowledge, which captures reusable signals from previously strong repositories, failure knowledge, which records unsuccessful outcomes and their diagnostic signals, and a historical-best repository, which preserves the strongest result found so far and prevents regression. These components collectively transform repeated repository generation into a persistent, knowledge-driven optimization process. We evaluate \approach using four frontier LLMs on two representative repository-level code generation benchmarks. Extensive experimental results demonstrate the effectiveness and efficiency of \approach, improving the functional score by up to 22.94 percentage points, increasing repository reuse to 81.58\%, and reducing cost by up to 53.63\% on RAL-Bench while maintaining broadly stable non-functional quality.

\end{abstract}




\maketitle

\section{Introduction}

Recent advances in large language models (LLMs) have substantially improved code generation from natural language descriptions \citep{hou2024large, guo2024deepseek, hui2024qwen2, bai2023qwen}.
Early research in this area primarily focused on function-level generation, where the goal is to produce a single function from a natural language description \citep{liu2023your}.
However, strong performance at the function level does not readily transfer to repository-level generation, where models must produce complete repositories with multiple files, modules, and dependencies \citep{zhang2024codeagent, li2024deveval, pan2026ral}.
As repository-level tasks grow in complexity, both human developers and LLMs face increasing difficulties in generating code that satisfies all requirements, especially in the presence of multi-file structure and intricate inter-file dependencies \citep{jin2024llms}. 
% \textbf{In such a setting, solving the same problem may require multiple attempts rather than a single successful generation.} 
In such settings, a single generation is often insufficient, and success may depend on repeated sampling, multiple trajectories, or more than one end-to-end attempt.
% For example, Codex improves from 28.8\% when only one sample is generated to 70.2\% when 100 samples are generated on HumanEval, and recent software engineering agent studies also benefit from multiple sampled trajectories \citep{chen2021evaluating, pan2024training}.
For example, on HumanEval, Codex scores 28.8\% when only one solution is sampled, but 70.2\% when 100 solutions are sampled \citep{chen2021evaluating}. Recent software engineering agent studies similarly benefit from multiple sampled trajectories \citep{pan2024training}.
% \hy{can you give some numbers and/or citations here. For example, how many rounds according to ...}
% However, most existing methods either refine a single end-to-end attempt or select among multiple sampled trajectories, rather than explicitly maintaining and reusing persistent task-specific state across separate attempts on the same task \citep{bi2024iterative, zhang2024codeagent}. \rw{revised, pls check}
% % However, existing methods are still mainly designed to optimize each attempt in isolation, therefore usually \hy{why usually, who else analyze cross-attempts}do not explicitly preserve and reuse task-specific signals across attempts. 
% As a result, later attempts may repeatedly explore ineffective directions or rediscover solutions that have already been partially obtained \citep{bi2024iterative}.

Given that a single generation is often insufficient, existing work has largely focused on iterative refinement and optimization of a single solving attempt. Some approaches revise generated code according to execution results \citep{madaan2023self, bi2024iterative}. Others optimize candidate solutions through search, mutation, evaluation, and selection \citep{novikov2025alphaevolve, hu2026controlled}. More recently, agent-based methods have been developed to support repository-level code generation through tool use, environment interaction, and execution-guided refinement \citep{zhang2024codeagent, xia2025live, lin2025se}. These advances have substantially improved code generation for complex tasks. 
\textbf{However, these methods still mainly refine a single end-to-end attempt or select among multiple sampled trajectories, rather than explicitly maintaining and reusing persistent task-specific state across separate attempts on the same task \citep{bi2024iterative, zhang2024codeagent}. In particular, they generally do not maintain accumulated success knowledge, accumulated failure knowledge, or the historical-best repository as reusable state across attempts.
As a result, across repeated attempts on the same task, subsequent attempts may repeatedly explore ineffective directions or rediscover solutions that have already been partially obtained \citep{bi2024iterative}.
}

In this paper, each attempt corresponds to one complete end-to-end solution of the same task, including all internal refinement, execution, and repair steps within that attempt. 
% Repeated repository-level code generation therefore requires more than within-attempt refinement and optimization\hy{the paper flow is not good here. It seems to be that it is back to the motivation/paragraph 1 again.}. It requires persistent task-specific state across attempts. \textbf{The key challenge lies not only in improving the current repository, but in preserving and reusing task-specific state accumulated from prior attempts.} 
% We therefore formulate repeated repository-level code generation around persistent task-specific state carried across attempts.
To address this limitation, we formulate repeated repository-level code generation around persistent task-specific state carried across attempts.
This state includes reusable success knowledge, reusable failure knowledge, and a preserved historical-best repository. 
% Existing methods generally do not explicitly maintain these forms of persistent state for the same problem, nor do they support direct reuse of a previously strong repository or preservation of the historical-best repository as a safeguard against regression. As a result, they may continue to incur redundant exploration and refinement costs even after useful state has already been accumulated.
Existing methods generally do not explicitly maintain these forms of persistent state for the same problem, nor do they support direct reuse of a previously strong repository or preservation of the historical-best repository as a safeguard against regression. \emph{This creates a key challenge:} even after useful state has been accumulated, later attempts may still incur redundant exploration and refinement costs.

To address this challenge, we propose \textbf{\approach}, a novel framework for repository-level code generation based on cross-attempt knowledge optimization. Rather than treating each attempt as an isolated optimization process, \approach maintains persistent task-specific state from prior attempts and uses it to guide subsequent decisions. Specifically, this state consists of success knowledge that captures reusable signals from previously strong repositories, failure knowledge that records unsuccessful outcomes and their diagnostic signals, and a historical-best repository that preserves the strongest result found so far and prevents regression in the final output. With these components, \approach can directly reuse a strong repository when appropriate, guide new generation with accumulated knowledge, and fall back to the historical-best repository when necessary. In this way, \approach transforms repeated repository-level generation from isolated attempts into a persistent, knowledge-driven optimization process.

We evaluate \approach on two representative benchmarks for repository-level code generation, RAL-Bench \citep{pan2026ral} and NL2Repo-Bench \citep{ding2025nl2repo}, using four frontier LLMs, including GPT-5-2025-08-07, DeepSeek-V3-0324, Claude-Sonnet-4.5-20250929, and Gemini-3-Pro-Preview. 
Experimental results show that \approach consistently outperforms strong baselines on repository-level code generation. On RAL-Bench, across different backbone models and repeated attempts on the same problem, \approach improves the functional score by up to 22.94 points, increases repository reuse to as high as 81.58\%, and reduces cost by up to 53.63\% from the first to the fourth attempt, while maintaining broadly stable non-functional quality. These results indicate that \approach increasingly reuses persistent task-specific state, reduces redundant refinement, and preserves the historical-best repository when necessary, thereby yielding more stable optimization behavior and favorable cost--effectiveness trade-offs.

In summary, the main contributions of this paper are as follows:
\begin{itemize}

    % \item We identify an overlooked challenge in repository-level code generation: existing methods mainly optimize within a single end-to-end attempt, but usually do not maintain persistent task-specific state across repeated attempts on the same problem.

    % \item We identify an overlooked challenge in repository-level code generation, where existing methods usually optimize each attempt in isolation, and we propose \approach, a novel framework for repository-level code generation based on cross-attempt knowledge optimization, which transforms repeated repository generation from isolated attempts into a persistent, knowledge-driven optimization process.
    \item We identify an overlooked challenge in repository-level code generation, as existing methods still optimize each attempt in isolation. To address this challenge, we propose \approach, a novel framework for repository-level code generation based on cross-attempt knowledge optimization. \approach transforms repeated repository generation from isolated attempts into a persistent, knowledge-driven optimization process.

    \item We introduce three key components of this persistent state in \approach: \cnum{myitem}{1} Success Knowledge for capturing reusable signals from previously strong repositories, \cnum{myitem}{2} Failure Knowledge for recording unsuccessful outcomes and their diagnostic signals, and \cnum{myitem}{3} the historical-best repository for preserving the strongest result found so far and preventing regression in the final output.

    \item We conduct extensive experiments on two representative benchmarks for repository-level code generation, RAL-Bench and NL2Repo-Bench, using multiple frontier LLMs. The results show that \approach consistently improves effectiveness and stability, reduces redundant refinement, and improves efficiency.
\end{itemize}

\section{Motivating Example}

\begin{figure}[t]
    \centering
    \includegraphics[width=\columnwidth]{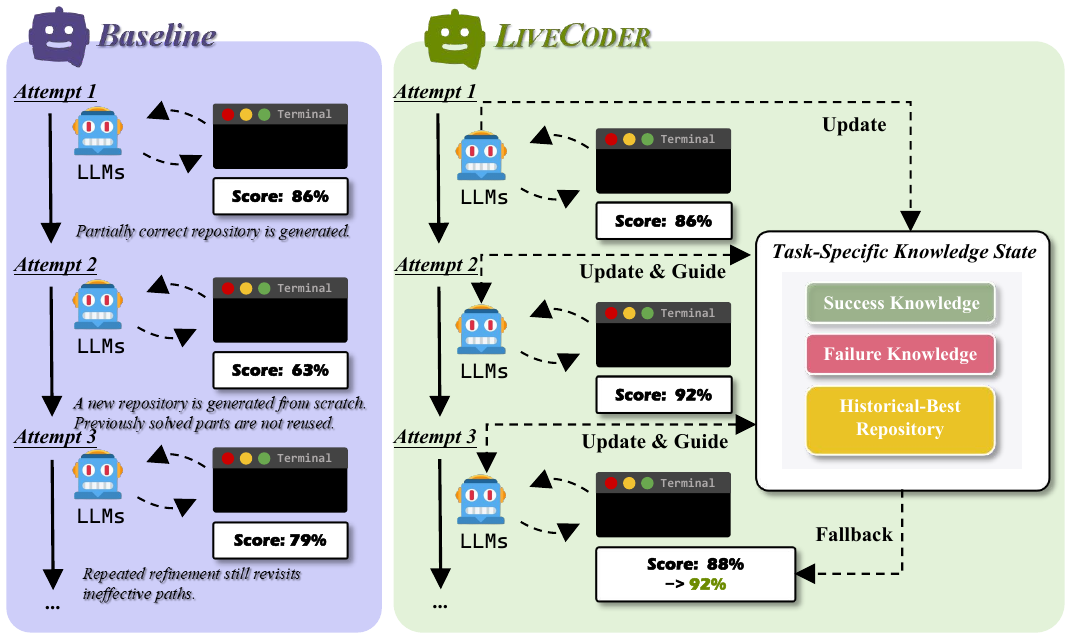}
    \caption{A motivating example of repeated repository-level code generation on the same problem. Existing methods may treat repeated attempts as isolated retries and regress below earlier stronger results, whereas \approach preserves task-specific state across attempts and protects the historical-best repository.}
    \label{fig:motivating_example}

    \vspace{-5px}
\end{figure}

Fig.~\ref{fig:motivating_example} shows a motivating example of repeated repository-level code generation on the same problem. In this example, an existing baseline achieves a functional score of 86\% in Attempt~1, but then drops to 63\% in Attempt~2 and recovers only to 79\% in Attempt~3. Although the first attempt already produces a partially correct repository, later attempts still fail to improve upon it in a stable manner. Instead, later attempts remain non-cumulative: previously solved parts are not retained, ineffective paths are revisited, and repository quality fluctuates across attempts.

This behavior reveals a key limitation of existing approaches. Although they may refine solutions within a single attempt, repeated attempts on the same problem are still largely treated as isolated retries. As a result, useful signals from prior strong attempts are not explicitly preserved or reused, previously exposed failures are not systematically carried forward, and later attempts may even regress below stronger earlier repositories.

This also differs from real software development, where prior effective solutions and exposed failure patterns are rarely discarded when the same task is revisited. In contrast, \approach explicitly preserves task-specific state across attempts, so that later attempts are guided rather than restarted from scratch. In the example, Attempt~2 improves the score from 86\% to 92\%, and the weaker output in Attempt~3 does not overwrite the stronger historical-best repository.

This example highlights the central intuition of our work. Repeated repository-level code generation should not be treated as repeated optimization from scratch, but as a persistent optimization process across attempts. The key is to preserve what has already worked, avoid what has already failed, and protect the strongest repository found so far.

\section{Framework}

\subsection{Overview}

\begin{figure*}[t]
    \centering
    \includegraphics[width=0.7\textwidth]{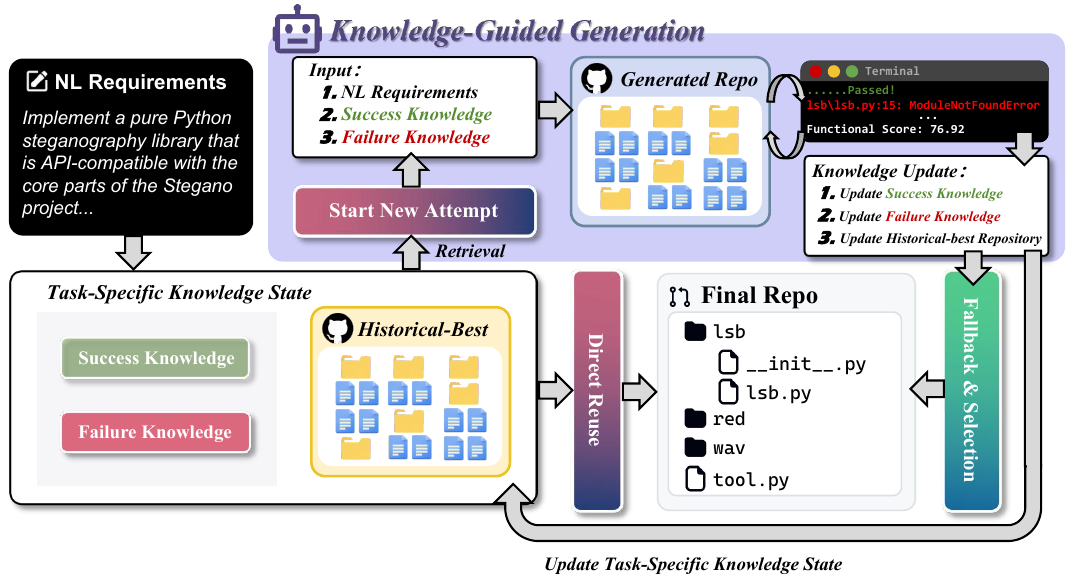}
    \caption{Overview of \approach. For each repository-level task, \approach maintains a persistent task-specific state consisting of Success Knowledge, Failure Knowledge, and the historical-best repository. This state guides subsequent attempts, while historical-best preservation prevents regression below the strongest repository found so far.}
    \label{fig:framework_overview}
    \vspace{-5px}
\end{figure*}

We formulate repository-level code generation as a repeated-attempt setting. Given a natural language requirement $x$, the goal is to generate a complete repository $R$ that satisfies the requirement. Unlike conventional settings that optimize a single attempt in isolation, we consider repeated attempts on the same task. In the $t$-th attempt, the framework generates a repository $R_t$ and obtains its functional score $s_t$. Across attempts, the framework maintains a persistent task-specific state consisting of Success Knowledge, Failure Knowledge, and the historical-best repository with its score $(R^{*}, s^{*})$. The objective is therefore not only to improve the repository generated in the current attempt, but also to preserve and reuse task-specific state across attempts.

Fig.~\ref{fig:framework_overview} illustrates the overall workflow of \approach, and Algorithm~\ref{alg:approach} presents the complete cross-attempt optimization procedure. Given a natural language requirement $x$, the framework first checks whether further generation is necessary. If the historical-best repository already achieves a full functional score, \approach directly reuses it and terminates. Otherwise, it launches a new attempt. Within each attempt, the generation module may perform multiple internal iterations of generation, execution, and repair before producing a candidate repository $R_t$. The resulting repository is then functionally evaluated to obtain its score $s_t$, and the persistent task-specific state is updated accordingly. High-scoring repositories contribute reusable positive signals to Success Knowledge, while lower-scoring ones contribute reusable failure signals to Failure Knowledge. At the end of each attempt, the framework compares the current repository $R_t$ with the historical-best repository. 
% If the current repository achieves a higher functional score ($s_t > s^{*}$), 
% % it updates the historical-best repository by setting $(R^{*}, s^{*}) \gets (R_t, s_t)$. 
% it becomes the new historical-best repository.
If the current repository achieves a higher functional score than the historical best ($s_t > s^{*}$), it becomes the new historical-best repository.
The historical-best repository is then used as the current output. If no repository reaches a full functional score after all allowed attempts, the framework returns the historical-best repository as the final output, ensuring that the final result never regresses below the best-performing repository found so far.

% \begin{algorithm}[t]
% \caption{Cross-Attempt Knowledge Optimization in \approach \yh{The specific value of $\tau_{\mathrm{succ}}$ and the rationale for choosing it do not appear to be provided later in the paper.}}
% \label{alg:approach}
% \begin{algorithmic}[1]
% \Require natural language requirement $x$, maximum number of attempts $A$, success threshold $\tau_{\mathrm{succ}}$, full functional score $s_{\mathrm{full}}$
% \Ensure final repository $R_{\mathrm{final}}$

% \State \textbf{Initialize:} Success Knowledge $K^{+} \gets \varnothing$, Failure Knowledge $K^{-} \gets \varnothing$
% \State \textbf{Initialize:} historical-best repository $R^{*} \gets \varnothing$, score $s^{*} \gets -\infty$

% \For{$t = 1$ to $A$}
%     \If{$R^{*} \neq \varnothing$ \textbf{and} $s^{*} = s_{\mathrm{full}}$}
%         \State \Return $R^{*}$
%     \EndIf

%     \State $R_t \gets \textsc{RunAttempt}(x, K^{+}, K^{-})$
%     \Comment{may include multiple internal iterations of generation, execution, and repair}

%     \State $s_t \gets \textsc{EvaluateFunctionalScore}(R_t)$

%     \If{$s_t \ge \tau_{\mathrm{succ}}$}
%         \State $K^{+} \gets \textsc{UpdateSuccessKnowledge}(K^{+}, R_t, s_t)$
%     \Else
%         \State $K^{-} \gets \textsc{UpdateFailureKnowledge}(K^{-}, R_t, s_t)$
%     \EndIf

%     \If{$s_t > s^{*}$}
%         \State $R^{*} \gets R_t$
%         \State $s^{*} \gets s_t$
%     \EndIf
% \EndFor

% \State $R_{\mathrm{final}} \gets R^{*}$
% \State \Return $R_{\mathrm{final}}$
% \end{algorithmic}
% \end{algorithm}

\begin{algorithm}[t]
\caption{Cross-Attempt Knowledge Optimization in \approach}
\label{alg:approach}
\begin{algorithmic}[1]
\Require natural language requirement $x$, maximum number of attempts $A$, full functional score $s_{\mathrm{full}}$
\Ensure final repository $R_{\mathrm{final}}$

\State \textbf{Initialize:} Success Knowledge $K^{+} \gets \varnothing$, Failure Knowledge $K^{-} \gets \varnothing$
\State \textbf{Initialize:} historical-best repository $R^{*} \gets \varnothing$, score $s^{*} \gets -\infty$

\For{$t = 1$ to $A$}
    \If{$R^{*} \neq \varnothing$ \textbf{and} $s^{*} \ge s_{\mathrm{full}}$}
        \State \Return $R^{*}$
    \EndIf

    \State $R_t \gets \textsc{RunAttempt}(x, K^{+}, K^{-})$
    \Comment{may include multiple internal iterations of generation, execution, and repair}

    \State $s_t \gets \textsc{EvaluateFunctionalScore}(R_t)$

    \State $K^{+} \gets \textsc{UpdateSuccessKnowledge}(K^{+}, R_t, s_t)$
    \State $K^{-} \gets \textsc{UpdateFailureKnowledge}(K^{-}, R_t, s_t)$

    \If{$s_t > s^{*}$}
        \State $R^{*} \gets R_t$
        \State $s^{*} \gets s_t$
    \EndIf
\EndFor

\State $R_{\mathrm{final}} \gets R^{*}$
\State \Return $R_{\mathrm{final}}$

\end{algorithmic}

\end{algorithm}

% \vspace{-10px}

\subsection{Task-Specific State Representation}

To support cross-attempt optimization, \approach maintains a persistent task-specific state for each problem. At attempt $t$, this state consists of Success Knowledge $K^{+}$, Failure Knowledge $K^{-}$, and the historical-best repository with its score $(R^{*}, s^{*})$. This state allows later attempts to benefit from reusable evidence accumulated from prior attempts.

\textit{\textbf{State representation.}}
Success Knowledge and Failure Knowledge are maintained as structured textual entries that can be directly injected into the generation module in subsequent attempts. Each Success Knowledge entry records reusable positive signals together with its source context, including the source attempt and the associated functional score. These signals may include effective repository structure, validated interface choices, successful dependency organization, and implementation patterns associated with strong functional outcomes. Each Failure Knowledge entry records reusable negative signals together with its source context, including the source attempt and the associated functional score. These signals may include missing repository components, broken inter-file dependencies, violated interface contracts, incomplete functional paths, and previously ineffective repair directions. The historical-best repository is stored separately from Success Knowledge and Failure Knowledge as a complete repository artifact, rather than being represented as a textual entry, because it serves direct reuse, fallback, and final selection.

\textit{\textbf{State update.}}
After functionally evaluating the repository generated in an attempt, \approach updates the persistent state based on the resulting functional score. In each attempt, the framework updates both Success Knowledge and Failure Knowledge rather than treating them as mutually exclusive alternatives. 
% \yh{There seems to be an inconsistency between Algorithm 1 and the description here. In Algorithm 1 (lines 9–13), Success Knowledge and Failure Knowledge are updated in a mutually exclusive manner based on $\tau_{\mathrm{succ}}$. However, it is explicitly stated here that "the framework updates both Success Knowledge and Failure Knowledge rather than treating them as mutually exclusive alternatives." Please reconcile these two descriptions to ensure consistency.}
Specifically, it extracts reusable positive signals from the repository and incorporates them into Success Knowledge, while also extracting reusable failure signals and incorporating them into Failure Knowledge. Independently, if the new repository achieves a higher functional score than the current historical-best repository, the framework replaces the historical-best repository with the new one. Through this update rule, the state preserves both reusable textual evidence and the best-performing repository observed so far.

\textit{\textbf{Signal extraction and structuring.}}
In each attempt, \approach uses an LLM-based extraction step to convert the generated repository and its execution feedback into structured textual entries. The framework first collects repository-level observations from the generated repository, including file organization, module interfaces, dependency usage, and implementation patterns. It also collects execution feedback, such as passed and failed tests, runtime errors, and the resulting functional score. Based on these inputs, the LLM summarizes reusable positive signals and reusable failure signals into a standardized schema with source context, associated score, and carry-over signals or constraints. In this way, Success Knowledge and Failure Knowledge are not written manually, but are automatically derived from the repository and its feedback through an explicit extraction and structuring process.

\textit{\textbf{State usage.}}
Before a new attempt begins, \approach retrieves the most relevant Success Knowledge and Failure Knowledge entries by semantic matching between the current natural language requirement and the stored textual entries. Specifically, the requirement and each state entry are encoded into embedding vectors, cosine similarity is used for ranking, and the top relevant entries are selected as guidance for repository generation. Success Knowledge is used as positive guidance that highlights promising repository-level decisions, whereas Failure Knowledge is used as negative guidance that discourages previously ineffective structures, dependencies, and repair directions. Conditioned on both types of signals, the generation module then produces a new repository, while the historical-best repository is retained as the best-so-far result across attempts.

\subsection{Success Knowledge}

The Success Knowledge module preserves reusable positive signals extracted from prior strong attempts on the same problem and uses them to steer later attempts toward already validated repository-level decisions. Its purpose is not to store the strongest repository itself, but to retain the positive signals that remain useful when later attempts continue generation rather than directly reusing the historical-best repository.

\begin{figure}[t]
\centering
\includegraphics[width=0.9\columnwidth]{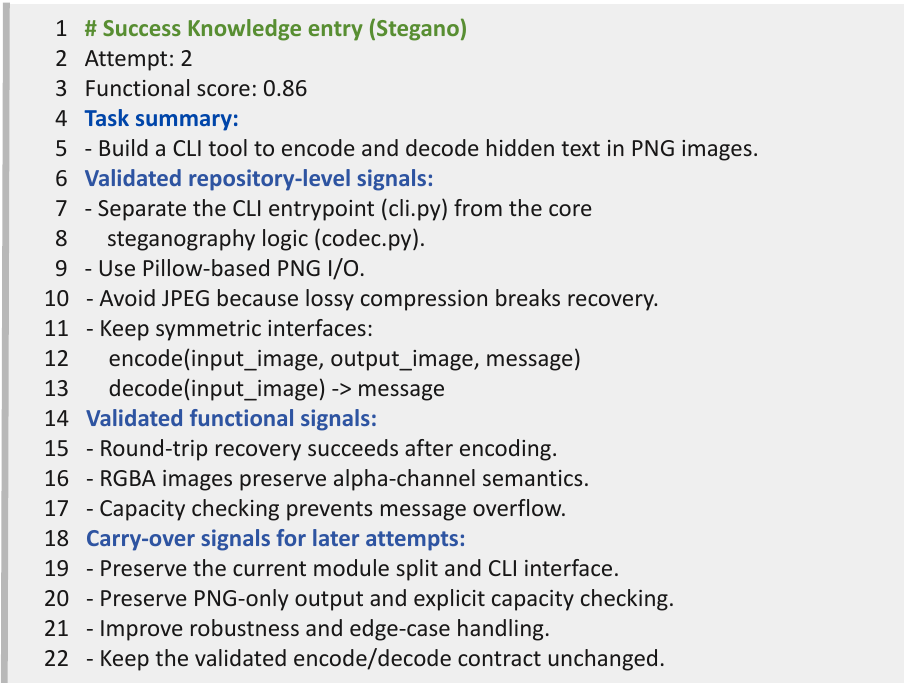}
\caption{Example Success Knowledge entry for a Stegano task.}
\label{fig:success_knowledge_stegano}
\vspace{-10px}
\end{figure}

\textit{\textbf{Role.}}
In repeated repository-level code generation, earlier attempts may already establish useful repository-level signals, such as effective multi-file organization, sound interface design, validated dependency choices, and functionally reliable implementation patterns. Discarding such signals and restarting each attempt from scratch can lead to redundant refinement and unstable optimization. Success Knowledge therefore preserves reusable positive signals across attempts, allowing later attempts to build on what has already been shown to work.

% \lh{comments about the details. I have written the specific content in Chinese in the review panel of Overleaf.} \rw{add the Signal extraction and structuring.}
\textit{\textbf{An Example.}}
Success Knowledge is represented as structured textual entries that summarize validated positive signals from prior strong attempts. Fig.~\ref{fig:success_knowledge_stegano} shows an example entry for a Stegano task. Rather than storing the entire repository, the entry records repository-level signals, functionally validated signals, and carry-over signals that can be directly injected into later generation. In this way, Success Knowledge captures what should be preserved from a strong attempt, while the historical-best repository is maintained separately as the strongest repository itself.

\subsection{Failure Knowledge}

\begin{figure}[t]
\centering
\includegraphics[width=0.9\columnwidth]{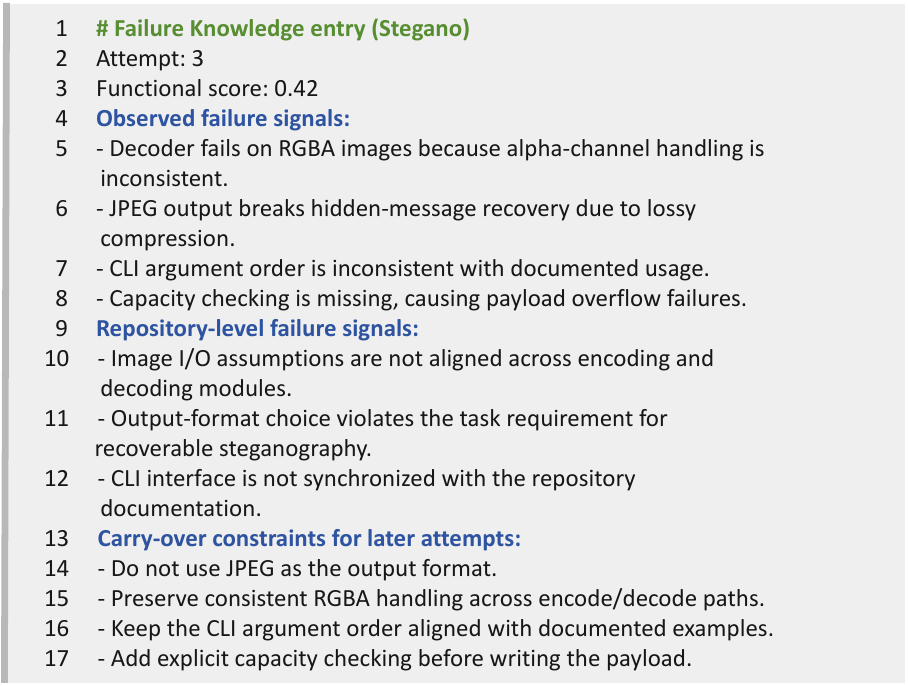}
\caption{Example Failure Knowledge entry for a Stegano task.}
\label{fig:failure_knowledge_stegano}
\vspace{-10px}
\end{figure}

The Failure Knowledge module preserves reusable negative signals extracted from prior weak attempts on the same problem and uses them to constrain later attempts from revisiting already ineffective repository-level directions. Its purpose is not to reject all prior outcomes, but to retain failure signals that remain informative for later generation.

\textit{\textbf{Role.}}
In repeated repository-level code generation, weak attempts may still reveal useful failure signals about why a generated repository does not satisfy the requirement. Such signals may include unmet functional requirements, missing or incomplete repository components, inconsistent inter-file dependencies, violated interface contracts, and previously ineffective repair directions. Discarding these signals and restarting each attempt from scratch can lead to repeated exploration of repository-level directions that have already been shown ineffective. Failure Knowledge therefore carries forward reusable negative signals across attempts, allowing later attempts to avoid known failure modes rather than rediscovering them.

\textit{\textbf{An Example.}}
Failure Knowledge is represented as structured textual entries that summarize reusable negative signals from prior weak attempts. Fig.~\ref{fig:failure_knowledge_stegano} shows an example entry for a Stegano task. Rather than storing the entire failed repository, each entry records observed failure signals, repository-level failure signals, and actionable carry-over constraints for later attempts. In this way, Failure Knowledge captures what should be avoided in future generation.

\subsection{Historical-Best Repository Preservation and Selection}

The historical-best repository preserves the best-performing repository achieved so far across repeated attempts and supports both direct reuse and final selection. Its purpose is to prevent later attempts from overwriting stronger earlier results and to ensure that the final output is always selected from the historical-best repository rather than the most recently generated repository.

\textit{\textbf{Role.}}
In repeated repository-level code generation, later attempts do not necessarily lead to consistent improvements. A newly generated repository may regress even when an earlier attempt has already solved part of the task more effectively. The historical-best repository therefore stabilizes optimization across attempts by preserving the best-performing repository found so far.

\textit{\textbf{Update.}}
After each attempt is completed and the resulting repository is functionally evaluated, \approach compares its functional score with that of the current historical-best repository. If the new repository achieves a higher functional score, it replaces the historical-best repository; otherwise, the historical-best repository is retained. Through this rule, \approach monotonically preserves the best-performing repository achieved so far across attempts.

\textit{\textbf{Selection and reuse.}}
Before a new attempt begins, if the historical-best repository already achieves a full functional score, \approach directly reuses it and terminates without further generation. Otherwise, the framework launches a new attempt while retaining the historical-best repository for later selection. After all allowed attempts are completed, the historical-best repository is returned as the final output. In this way, final selection is always performed over the best-performing repository retained across attempts, rather than defaulting to the most recently generated repository.

\section{Experiments and Results}

\begin{table*}[t]
\centering
\caption{\textbf{Comparison of self-evolving baselines under multiple backbone models on RAL-Bench.}
Numbers are percentages. Gray numbers in parentheses indicate the absolute change (percentage points) relative to \textbf{Direct} under the same backbone.}
\label{tab:benchmark_stats}
\small
\renewcommand{\arraystretch}{1.18}
\definecolor{DeepBlue}{RGB}{77,88,234}
\definecolor{LightBlue}{RGB}{235,241,250}
\renewcommand{\chg}[1]{\textcolor{gray}{\scriptsize\,({#1})}}

\resizebox{0.77\textwidth}{!}{%
\begin{tabular}{l|ccccccc}
\toprule
\textbf{Method} & \textbf{Functional} & \textbf{Non-functional} & \textbf{Maintainability} & \textbf{Security} & \textbf{Robustness} & \textbf{Efficiency} & \textbf{Resource} \\
\midrule

\multicolumn{8}{c}{\textcolor{DeepBlue}{\textbf{\textit{GPT-5-2025-08-07}}}} \\
\midrule
\textbf{Direct}            & 38.50 & 57.51 & 33.84 & 94.74 & 77.02 & 41.81 & 43.73 \\
\textbf{Self-Reflection}   & 44.14\chg{+5.64} & 55.02\chg{-2.49} & 32.11\chg{-1.73} & 97.37\chg{+2.63} & 78.69\chg{+1.67} & 42.11\chg{+0.30} & 42.93\chg{-0.80} \\
\textbf{SE-Agent}          & 44.65\chg{+6.15} & 54.46\chg{-3.05} & 37.24\chg{+3.40} & 97.37\chg{+2.63} & 77.69\chg{+0.67} & 26.32\chg{-15.49} & 38.43\chg{-5.30} \\
\textbf{AlphaEvolve}       & 43.70\chg{+5.20} & 59.60\chg{+2.09} & 35.99\chg{+2.15} & 100.00\chg{+5.26} & 82.99\chg{+5.97} & 39.62\chg{-2.19} & 38.43\chg{-5.30} \\
\textbf{CSE}               & 34.33\chg{-4.17} & 53.26\chg{-4.25} & 31.97\chg{-1.87} & 97.37\chg{+2.63} & 71.87\chg{-5.15} & 39.47\chg{-2.34} & 17.88\chg{-25.85} \\
\textbf{Live-SWE-Agent}    & 56.45\chg{+17.95} & 63.13\chg{+5.62} & 20.52\chg{-13.32} & 100.00\chg{+5.26} & 92.63\chg{+15.61} & 73.69\chg{+31.88} & 67.32\chg{+23.59} \\
\cellcolor{LightBlue}\textcolor{DeepBlue}{\textbf{Ours}}
& \cellcolor{LightBlue}\textcolor{DeepBlue}{\textbf{58.20}}\chg{+19.70}
& \cellcolor{LightBlue}\textcolor{DeepBlue}{\textbf{65.33}}\chg{+7.82}
& \cellcolor{LightBlue}\textcolor{DeepBlue}{\textbf{27.86}}\chg{-5.98}
& \cellcolor{LightBlue}\textcolor{DeepBlue}{\textbf{100.00}}\chg{+5.26}
& \cellcolor{LightBlue}\textcolor{DeepBlue}{\textbf{93.16}}\chg{+16.14}
& \cellcolor{LightBlue}\textcolor{DeepBlue}{\textbf{71.36}}\chg{+29.55}
& \cellcolor{LightBlue}\textcolor{DeepBlue}{\textbf{65.27}}\chg{+21.54} \\
\midrule

\multicolumn{8}{c}{\textcolor{DeepBlue}{\textbf{\textit{DeepSeek-V3-0324}}}} \\
\midrule
\textbf{Direct}            & 24.37 & 46.34 & 32.29 & 81.58 & 65.89 & 22.90 & 15.36 \\
\textbf{Self-Reflection}   & 26.30\chg{+1.93} & 50.43\chg{+4.09} & 37.32\chg{+5.03} & 97.37\chg{+15.79} & 72.25\chg{+6.36} & 23.68\chg{+0.78} & 17.54\chg{+2.18} \\
\textbf{SE-Agent}          & 25.18\chg{+0.81} & 52.53\chg{+6.19} & 39.04\chg{+6.75} & 97.37\chg{+15.79} & 70.62\chg{+4.73} & 18.42\chg{-4.48} & 15.96\chg{+0.60} \\
\textbf{AlphaEvolve}       & 26.78\chg{+2.41} & 55.75\chg{+9.41} & 38.88\chg{+6.59} & 100.00\chg{+18.42} & 77.46\chg{+11.57} & 28.95\chg{+6.05} & 15.73\chg{+0.37} \\
\textbf{CSE}               & 23.99\chg{-0.38} & 55.24\chg{+8.90} & 38.72\chg{+6.43} & 100.00\chg{+18.42} & 82.53\chg{+16.64} & 21.05\chg{-1.85} & 13.10\chg{-2.26} \\
\textbf{Live-SWE-Agent}    & 32.57\chg{+8.20} & 60.80\chg{+14.46} & 39.72\chg{+7.43} & 100.00\chg{+18.42} & 83.19\chg{+17.30} & 43.24\chg{+20.34} & 33.33\chg{+17.97} \\
\cellcolor{LightBlue}\textcolor{DeepBlue}{\textbf{Ours}}
& \cellcolor{LightBlue}\textcolor{DeepBlue}{\textbf{34.37}}\chg{+10.00}
& \cellcolor{LightBlue}\textcolor{DeepBlue}{\textbf{44.24}}\chg{-2.10}
& \cellcolor{LightBlue}\textcolor{DeepBlue}{\textbf{18.22}}\chg{-14.07}
& \cellcolor{LightBlue}\textcolor{DeepBlue}{\textbf{78.94}}\chg{-2.64}
& \cellcolor{LightBlue}\textcolor{DeepBlue}{\textbf{65.87}}\chg{-0.02}
& \cellcolor{LightBlue}\textcolor{DeepBlue}{\textbf{36.84}}\chg{+13.94}
& \cellcolor{LightBlue}\textcolor{DeepBlue}{\textbf{31.45}}\chg{+16.09} \\
\midrule

\multicolumn{8}{c}{\textcolor{DeepBlue}{\textbf{\textit{Claude-Sonnet-4.5-20250929}}}} \\
\midrule
\textbf{Direct}            & 39.32 & 49.64 & 30.19 & 84.21 & 66.93 & 31.50 & 33.94 \\
\textbf{Self-Reflection}   & 44.74\chg{+5.42} & 53.98\chg{+4.34} & 36.21\chg{+6.02} & 97.37\chg{+13.16} & 76.67\chg{+9.74} & 23.68\chg{-7.82} & 39.42\chg{+5.48} \\
\textbf{SE-Agent}          & 43.73\chg{+4.41} & 53.80\chg{+4.16} & 36.30\chg{+6.11} & 97.37\chg{+13.16} & 76.99\chg{+10.06} & 23.68\chg{-7.82} & 42.50\chg{+8.56} \\
\textbf{AlphaEvolve}       & 42.26\chg{+2.94} & 58.61\chg{+8.97} & 35.91\chg{+5.72} & 100.00\chg{+15.79} & 82.29\chg{+15.36} & 39.47\chg{+7.97} & 31.50\chg{-2.44} \\
\textbf{CSE}               & 37.01\chg{-2.31} & 53.70\chg{+4.06} & 29.69\chg{-0.50} & 96.05\chg{+11.84} & 67.27\chg{+0.34} & 42.10\chg{+10.60} & 33.98\chg{+0.04} \\
\textbf{Live-SWE-Agent}    & 52.10\chg{+12.78} & 59.78\chg{+10.14} & 29.24\chg{-0.95} & 100.00\chg{+15.79} & 85.49\chg{+18.56} & 49.68\chg{+18.18} & 46.79\chg{+12.85} \\
\cellcolor{LightBlue}\textcolor{DeepBlue}{\textbf{Ours}}
& \cellcolor{LightBlue}\textcolor{DeepBlue}{\textbf{67.00}}\chg{+27.68}
& \cellcolor{LightBlue}\textcolor{DeepBlue}{\textbf{64.81}}\chg{+15.17}
& \cellcolor{LightBlue}\textcolor{DeepBlue}{\textbf{34.16}}\chg{+3.97}
& \cellcolor{LightBlue}\textcolor{DeepBlue}{\textbf{100.00}}\chg{+15.79}
& \cellcolor{LightBlue}\textcolor{DeepBlue}{\textbf{80.26}}\chg{+13.33}
& \cellcolor{LightBlue}\textcolor{DeepBlue}{\textbf{65.78}}\chg{+34.28}
& \cellcolor{LightBlue}\textcolor{DeepBlue}{\textbf{64.78}}\chg{+30.84} \\
\midrule

\multicolumn{8}{c}{\textcolor{DeepBlue}{\textbf{\textit{Gemini-3-Pro-Preview}}}} \\
\midrule
\textbf{Direct}            & 43.86 & 50.57 & 31.50 & 76.75 & 66.36 & 43.69 & 41.23 \\
\textbf{Self-Reflection}   & 46.53\chg{+2.67} & 55.44\chg{+4.87} & 38.09\chg{+6.59} & 97.37\chg{+20.62} & 80.18\chg{+13.82} & 41.93\chg{-1.76} & 41.61\chg{+0.38} \\
\textbf{SE-Agent}          & 44.55\chg{+0.69} & 55.97\chg{+5.40} & 37.24\chg{+5.74} & 97.37\chg{+20.62} & 79.04\chg{+12.68} & 23.64\chg{-20.05} & 41.19\chg{-0.04} \\
\textbf{AlphaEvolve}       & 49.55\chg{+5.69} & 59.94\chg{+9.37} & 37.63\chg{+6.13} & 97.79\chg{+21.04} & 80.93\chg{+14.57} & 47.89\chg{+4.20} & 38.63\chg{-2.60} \\
\textbf{CSE}               & 31.76\chg{-12.10} & 56.07\chg{+5.50} & 38.31\chg{+6.81} & 97.36\chg{+20.61} & 72.46\chg{+6.10} & 31.57\chg{-12.12} & 29.37\chg{-11.86} \\
\textbf{Live-SWE-Agent}    & 58.76\chg{+14.90} & 64.34\chg{+13.77} & 35.87\chg{+4.37} & 97.46\chg{+20.71} & 79.54\chg{+13.18} & 50.34\chg{+6.65} & 51.32\chg{+10.09} \\
\cellcolor{LightBlue}\textcolor{DeepBlue}{\textbf{Ours}}
& \cellcolor{LightBlue}\textcolor{DeepBlue}{\textbf{69.16}}\chg{+25.30}
& \cellcolor{LightBlue}\textcolor{DeepBlue}{\textbf{67.03}}\chg{+16.46}
& \cellcolor{LightBlue}\textcolor{DeepBlue}{\textbf{36.78}}\chg{+5.28}
& \cellcolor{LightBlue}\textcolor{DeepBlue}{\textbf{98.68}}\chg{+21.93}
& \cellcolor{LightBlue}\textcolor{DeepBlue}{\textbf{86.84}}\chg{+20.48}
& \cellcolor{LightBlue}\textcolor{DeepBlue}{\textbf{71.25}}\chg{+27.56}
& \cellcolor{LightBlue}\textcolor{DeepBlue}{\textbf{64.02}}\chg{+22.79} \\
\bottomrule
\end{tabular}%
}
\end{table*}

We evaluate \approach through the following research questions (RQs):

\begin{itemize}
    \item \textbf{RQ1. Can \approach improve repository-level code generation compared with strong baselines?} We compare \approach with representative approaches for direct generation, iterative refinement, and agentic code generation across multiple backbone LLMs and benchmarks.

    \item \textbf{RQ2. Can persistent cross-attempt knowledge lead to sustained improvement over repeated attempts?} We investigate whether accumulated knowledge across repeated attempts on the same task can improve generation quality and reduce redundant exploration, compared with treating each attempt independently.

    \item \textbf{RQ3. Which mechanisms are most critical to the effectiveness of \approach?} We conduct ablation studies to isolate the contributions of Success Knowledge, Failure Knowledge, and preservation of the historical-best repository within the overall framework.

    \item \textbf{RQ4. What do cost and error analyses reveal about the behavior of \approach?} We analyze how the computational cost of \approach evolves across repeated attempts and which failure patterns remain in unsuccessful cases.
\end{itemize}

\subsection{Experiment Settings}

\textbf{\textit{Benchmarks.} }
Following prior work on repository-level code generation, we conduct experiments on two representative benchmarks, RAL-Bench \citep{pan2026ral} and NL2Repo-Bench \citep{ding2025nl2repo}. 
RAL-Bench focuses on generating complete repositories from high-level requirements in realistic application settings, requiring models to coordinate multiple files, modules, interfaces, and execution dependencies. 
In contrast, NL2Repo-Bench evaluates natural-language-to-repository generation and emphasizes the ability to translate user requirements into coherent repository implementations with appropriate multi-file organization and inter-file coordination. Together, these two benchmarks provide complementary evaluation settings for repository-level code generation and enable us to assess the effectiveness of persistent cross-attempt optimization across different task distributions.

\textbf{\textit{Metrics.}}
We evaluate generated repositories with functional correctness as the primary criterion and non-functional quality as a supplementary perspective. Functional correctness is measured by the \textit{functional test pass rate}. This metric also determines the preservation of the historical-best repository and the triggering of fallback in \approach.  
% However, in repository-level code generation, passing tests alone does not fully capture the practical value of a generated repository, especially when different approaches may produce functionally similar but qualitatively different outcomes across repeated attempts. Therefore, following the ISO/IEC~25010 quality model \citep{estdale2018applying}, we further assess five non-functional dimensions that are critical to real-world software development: \textit{maintainability, security, robustness, efficiency, and resource usage}. These dimensions are normalized and aggregated into a single score using AHP-derived weights. Detailed definitions of the metrics, normalization formulas, and the AHP derivation are provided in \href{https://anonymous.4open.science/r/LiveCoder}
%      {\textbf{\textcolor{oai-orange}{the anonymous repository}}}.
However, in repository-level code generation, functional correctness alone does not fully capture repository quality. Therefore, following the ISO/IEC~25010 quality model \citep{estdale2018applying}, we additionally evaluate five non-functional dimensions: \emph{maintainability}, \emph{security}, \emph{robustness}, \emph{efficiency}, and \emph{resource usage}. 
% For maintainability, we compute the Maintainability Index (MI) using the Coleman--Oman formulation \citep{coleman1994mi}:
% {\setlength{\abovedisplayskip}{2pt}
% \setlength{\belowdisplayskip}{2pt}
% \[
% MI=\max\!\left(0,\frac{171-5.2\ln V-0.23G-16.2\ln L}{171}\times 100\right),
% \]
% }
% where $V$, $G$, and $L$ denote Halstead volume, cyclomatic complexity, and lines of code, respectively. 
The lower-bound clipping at 0 follows the widely used Visual Studio code-metrics definition of MI \citep{microsoft2025maintainabilityindex}. For security, robustness, efficiency, and resource usage, we directly use the official benchmark evaluators from RAL-Bench and NL2Repo-Bench \citep{pan2026ral,ding2025nl2repo}, and then normalize dimension scores to a common scale. We aggregate the five normalized dimensions into a single non-functional score using AHP-derived weights:
{\setlength{\abovedisplayskip}{2pt}
\setlength{\belowdisplayskip}{2pt}
\[
S_{\mathrm{nonfunc}}=\sum_{k=1}^{5} w_k \hat{s}_k,
\]
}
where $\hat{s}_k$ is the normalized score of the $k$-th dimension and $w_k$ is its corresponding AHP-derived weight. 
% Detailed metric definitions (including MI computation), evaluator scripts, normalization formulas, and AHP weight derivation are provided in \href{https://anonymous.4open.science/r/LiveCoder}{\textbf{\textcolor{oai-orange}{the anonymous repository}}}.
Detailed scoring rules, normalization formulas, and weight derivation are provided in our repository.

\textbf{\textit{Comparative Methods.}}
We compare \approach with representative baselines spanning one-shot generation, iterative self-improvement, evolutionary code search, and repository-level interactive software agents. Together, these baselines cover the main categories of alternatives in our setting, including methods that generate a repository in a single pass, iteratively refine solutions within an attempt, or improve code through interaction with execution environments.

\begin{itemize}
    \item \textbf{Direct prompting \citep{pan2026ral}} prompts the LLM with the original task requirement to generate the target repository in a single pass, without iterative refinement, external feedback, or cross-attempt reuse.

    \item \textbf{Self-Reflection \citep{madaan2023self}} iteratively improves generation by producing natural-language reflections on previous outputs, which are then used to guide subsequent iterations.

    \item \textbf{SE-Agent \citep{lin2025se}} is a self-evolution framework that improves software engineering performance by revising, recombining, and refining prior reasoning trajectories across iterations.

    \item \textbf{AlphaEvolve \citep{novikov2025alphaevolve}} is an evolutionary code generation framework that iteratively mutates candidate programs and selects stronger ones based on automated evaluation feedback.

    \item \textbf{Controlled Self-Evolution (CSE) \citep{hu2026controlled}} improves code generation through collaborative self-evolution, in which multiple candidate solutions are iteratively generated, evaluated, and refined within a shared solving process.

    \item \textbf{Live-SWE-Agent \citep{xia2025live}} is a repository-level software engineering agent that continuously interacts with a live repository and uses execution feedback from tests and tools to modify and repair code.
\end{itemize}

\textbf{\textit{Implementation details.}}
We evaluate \approach with four frontier LLMs, including GPT-5-2025-08-07 \citep{singh2025openai}, DeepSeek-V3-0324 \citep{liu2024deepseek}, Claude-Sonnet-4.5-20250929 \citep{anthropic2025claudesonnet45}, and Gemini-3-Pro-Preview \citep{google2025gemini3propreview}. 
At the time of evaluation, we accessed these models through their official APIs when available, or through officially released model artifacts and compatible serving interfaces otherwise.
Unless otherwise specified, all approaches use the default context window of each model and greedy decoding with temperature set to 0. The maximum number of attempts is set to 4 for \approach.
% \lh{comments about semantic similarity.}
% \lh{comments about details of the threshold.}

For comparative methods, we reproduce them with released code or available official prompts. For iterative baselines, we use a maximum of 4 iterations as the refinement budget, following prior work on iterative code generation and refinement \citep{madaan2023self}. All reproduced methods are evaluated under the same execution environment.

To ensure a fair comparison, all methods are given the same problem descriptions as input for repository generation. The generated repositories are then executed and evaluated in a unified Python~3.9 environment, with a per-execution timeout of 300 seconds. This setup ensures that all approaches receive identical external feedback under the same execution conditions, thereby enabling fair and controlled comparisons.

\subsection{RQ1: Accuracy Comparison}

\begin{table}[t]
\centering
\caption{\textbf{Comparison of self-evolving baselines on NL2Repo-Bench across backbone models.}
All numbers are percentages. Gray numbers in parentheses indicate the absolute change relative to \textbf{Direct} in the same setting. Bold numbers indicate the best baseline in each column. Blue-highlighted rows denote \textbf{Ours}, and red-highlighted \textbf{Imp.} rows show its relative change vs. the best baseline.}
\label{tab:rq1_nl2repobench}
\small
\renewcommand{\arraystretch}{1.12}
\setlength{\tabcolsep}{4.3pt}

\definecolor{TblBlue}{RGB}{77,88,234}
\definecolor{TblBlueFill}{RGB}{235,241,250}
\definecolor{TblRed}{RGB}{196,59,59}
\definecolor{TblRedFill}{RGB}{252,239,239}

\newcommand{\tblbest}[1]{\textbf{#1}}
\newcommand{\tblours}[1]{\textcolor{TblBlue}{\textbf{#1}}}
\newcommand{\tblimppos}[1]{\textcolor{TblRed}{\textbf{#1}}}
\newcommand{\tblimpneg}[1]{\textcolor{gray}{\textbf{#1}}}

\resizebox{\columnwidth}{!}{%
\begin{tabular}{lcccc}
\toprule
\multirow{2}{*}{\textbf{Method}} & \textbf{Overall} & \textbf{Easy} & \textbf{Medium} & \textbf{Hard} \\
& \textbf{Score (\%)} & ($\le$1.5k LOC) & (1.5k--4k LOC) & ($\ge$4k LOC) \\
\midrule

\multicolumn{5}{c}{\textcolor{TblBlue}{\textbf{\textit{GPT-5-2025-08-07}}}} \\
\midrule
\textbf{Direct}            & 21.7\chg{+0.0}               & 38.4\chg{+0.0}               & 20.7\chg{+0.0}               & 9.6\chg{+0.0}               \\
\textbf{Self-Reflection}   & 24.7\chg{+3.0}               & 40.9\chg{+2.5}               & 24.5\chg{+3.8}               & 12.1\chg{+2.5}              \\
\textbf{SE-Agent}          & \tblbest{25.4}\chg{+3.7}     & \tblbest{44.5}\chg{+6.1}     & 23.6\chg{+2.9}               & \tblbest{12.6}\chg{+3.0}    \\
\textbf{AlphaEvolve}       & 25.1\chg{+3.4}               & 41.8\chg{+3.4}               & 25.1\chg{+4.4}               & 11.7\chg{+2.1}              \\
\textbf{CSE}               & 19.7\chg{-2.0}               & 41.1\chg{+2.7}               & 15.8\chg{-4.9}               & 7.9\chg{-1.7}               \\
\textbf{Live-SWE-Agent}    & 24.9\chg{+3.2}               & 35.8\chg{-2.6}               & \tblbest{30.4}\chg{+9.7}     & 8.1\chg{-1.5}               \\
\rowcolor{TblBlueFill}
\textbf{Ours}              & \tblours{27.6}\chg{+5.9}     & \tblours{30.2}\chg{-8.2}     & \tblours{34.5}\chg{+13.8}    & \tblours{5.5}\chg{-4.1}     \\
\rowcolor{TblRedFill}
\textcolor{TblRed}{\textbf{Imp.}}
                          & \tblimppos{+8.7\%}           & \tblimpneg{-32.1\%}          & \tblimppos{+13.5\%}          & \tblimpneg{-56.3\%}         \\
\midrule

\multicolumn{5}{c}{\textcolor{TblBlue}{\textbf{\textit{DeepSeek-V3-0324}}}} \\
\midrule
\textbf{Direct}            & 22.2\chg{+0.0}               & 35.7\chg{+0.0}               & 24.6\chg{+0.0}               & 12.1\chg{+0.0}              \\
\textbf{Self-Reflection}   & \tblbest{26.3}\chg{+4.1}     & 40.6\chg{+4.9}               & 26.3\chg{+1.7}               & 14.9\chg{+2.8}              \\
\textbf{SE-Agent}          & 24.1\chg{+1.9}               & 39.5\chg{+3.8}               & 23.4\chg{-1.2}               & \tblbest{15.3}\chg{+3.2}    \\
\textbf{AlphaEvolve}       & 24.7\chg{+2.5}               & \tblbest{44.2}\chg{+8.5}     & 30.1\chg{+5.5}               & 13.7\chg{+1.6}              \\
\textbf{CSE}               & 22.1\chg{-0.1}               & 40.1\chg{+4.4}               & 20.8\chg{-3.8}               & 9.4\chg{-2.7}               \\
\textbf{Live-SWE-Agent}    & 24.2\chg{+2.0}               & 36.1\chg{+0.4}               & \tblbest{30.4}\chg{+5.8}     & 5.7\chg{-6.4}               \\
\rowcolor{TblBlueFill}
\textbf{Ours}              & \tblours{26.7}\chg{+4.5}     & \tblours{30.3}\chg{-5.4}     & \tblours{31.2}\chg{+6.6}     & \tblours{17.9}\chg{+5.8}    \\
\rowcolor{TblRedFill}
\textcolor{TblRed}{\textbf{Imp.}}
                          & \tblimppos{+1.5\%}           & \tblimpneg{-31.4\%}          & \tblimppos{+2.6\%}           & \tblimppos{+17.0\%}         \\
\midrule

\multicolumn{5}{c}{\textcolor{TblBlue}{\textbf{\textit{Claude-Sonnet-4.5-20250929}}}} \\
\midrule
\textbf{Direct}            & 39.9\chg{+0.0}               & 55.3\chg{+0.0}               & 43.0\chg{+0.0}               & 21.4\chg{+0.0}              \\
\textbf{Self-Reflection}   & \tblbest{44.8}\chg{+4.9}     & 60.0\chg{+4.7}               & \tblbest{49.3}\chg{+6.3}     & \tblbest{26.1}\chg{+4.7}    \\
\textbf{SE-Agent}          & 44.4\chg{+4.5}               & \tblbest{62.3}\chg{+7.0}     & 47.7\chg{+4.7}               & 25.3\chg{+3.9}              \\
\textbf{AlphaEvolve}       & 43.1\chg{+3.2}               & 58.3\chg{+3.0}               & 47.3\chg{+4.3}               & 24.7\chg{+3.3}              \\
\textbf{CSE}               & 38.2\chg{-1.7}               & 61.2\chg{+5.9}               & 40.1\chg{-2.9}               & 16.9\chg{-4.5}              \\
\textbf{Live-SWE-Agent}    & 41.9\chg{+2.0}               & 57.1\chg{+1.8}               & 45.9\chg{+2.9}               & 23.9\chg{+2.5}              \\
\rowcolor{TblBlueFill}
\textbf{Ours}              & \tblours{45.8}\chg{+5.9}     & \tblours{60.9}\chg{+5.6}     & \tblours{51.1}\chg{+8.1}     & \tblours{27.5}\chg{+6.1}    \\
\rowcolor{TblRedFill}
\textcolor{TblRed}{\textbf{Imp.}}
                          & \tblimppos{+2.2\%}           & \tblimpneg{-2.2\%}           & \tblimppos{+3.7\%}           & \tblimppos{+5.4\%}          \\
\midrule

\multicolumn{5}{c}{\textcolor{TblBlue}{\textbf{\textit{Gemini-3-Pro-Preview}}}} \\
\midrule
\textbf{Direct}            & 34.2\chg{+0.0}               & 44.9\chg{+0.0}               & 40.9\chg{+0.0}               & 16.8\chg{+0.0}              \\
\textbf{Self-Reflection}   & \tblbest{38.1}\chg{+3.9}     & 50.2\chg{+5.3}               & 43.8\chg{+2.9}               & \tblbest{20.3}\chg{+3.5}    \\
\textbf{SE-Agent}          & 37.6\chg{+3.4}               & 48.3\chg{+3.4}               & \tblbest{44.2}\chg{+3.3}     & 19.4\chg{+2.6}              \\
\textbf{AlphaEvolve}       & 37.4\chg{+3.2}               & \tblbest{51.2}\chg{+6.3}     & 42.1\chg{+1.2}               & 19.7\chg{+2.9}              \\
\textbf{CSE}               & 33.3\chg{-0.9}               & 46.9\chg{+2.0}               & 38.2\chg{-2.7}               & 15.5\chg{-1.3}              \\
\textbf{Live-SWE-Agent}    & 34.4\chg{+0.2}               & 40.4\chg{-4.5}               & 42.8\chg{+1.9}               & 17.3\chg{+0.5}              \\
\rowcolor{TblBlueFill}
\textbf{Ours}              & \tblours{38.4}\chg{+4.2}     & \tblours{49.6}\chg{+4.7}     & \tblours{45.3}\chg{+4.4}     & \tblours{20.5}\chg{+3.7}    \\
\rowcolor{TblRedFill}
\textcolor{TblRed}{\textbf{Imp.}}
                          & \tblimppos{+0.8\%}           & \tblimpneg{-3.1\%}           & \tblimppos{+2.5\%}           & \tblimppos{+1.0\%}          \\
\bottomrule
\end{tabular}%
}
\end{table}

The comparison results are reported in Tables~\ref{tab:benchmark_stats} and~\ref{tab:rq1_nl2repobench}. Overall, \approach achieves the best overall score on NL2Repo-Bench and the best functional score on RAL-Bench under all four backbone models. Compared with \textbf{Direct}, \approach improves the overall score on NL2Repo-Bench by 5.9, 4.5, 5.9, and 4.2 points for GPT-5, DeepSeek-V3, Claude-Sonnet-4.5, and Gemini-3-Pro-Preview, respectively. On RAL-Bench, it improves the functional score by 19.70, 10.00, 27.68, and 25.30 points, respectively. These results show that the advantage of \approach is consistent across repository-level benchmarks and frontier backbone models.

The gains are not uniform across difficulty settings. On NL2Repo-Bench, \approach is most consistently effective in the \textbf{Medium} bucket, where it achieves the best result under all four backbones. This trend also helps explain the strong overall results, since the Medium bucket constitutes the largest portion of NL2Repo-Bench and therefore contributes most strongly to the overall score \citep{ding2025nl2repo}. By contrast, the \textbf{Easy} bucket leaves less room for improvement, while the \textbf{Hard} bucket remains challenging, especially for GPT-5. This suggests that cross-attempt knowledge optimization is particularly effective on repositories of moderate complexity, while very hard cases still expose substantial difficulty.

The RAL-Bench results further support this conclusion. \approach achieves the best functional performance under all four backbones, with especially large gains on stronger models such as Claude-Sonnet-4.5 and Gemini-3-Pro-Preview. At the same time, the non-functional results are more mixed, indicating that the main strength of \approach lies in improving repository-level executability, while non-functional optimization remains more model-dependent.

\begin{tcolorbox}[
  enhanced,
  colback=grey,                % 背景色
  colframe=teal!60!black,       % 边框颜色
  boxrule=0.6pt,                % 边框线宽
  arc=10pt,                     % 圆角大小
  left=4mm,right=4mm,           % 内边距
  top=2mm,bottom=2mm,
  drop shadow={black!40!white}, % 右下角阴影
]
\textbf{\textit{Answer to RQ1:} }
\textit{\approach achieves the best overall score on NL2Repo-Bench and the best functional score on RAL-Bench under all four frontier LLMs. This shows that cross-attempt knowledge optimization effectively improves repository-level generation by preserving validated decisions and reducing repeated ineffective exploration.}

\end{tcolorbox}

\subsection{RQ2: Impact of Knowledge Evolution Across Attempts}

\begin{table}[t]
\centering
\caption{Impact of knowledge evolution across attempts on RAL-Bench. 
We report the functional score (Func.), non-functional score (Non-func.), and repository reuse rate. Reuse (\%) denotes the percentage of tasks whose current attempt directly reuses the historical-best repository. Gray numbers in parentheses indicate the absolute change relative to Attempt~1 for the same model.}
\label{tab:rq2_knowledge_evolution}
\small
\renewcommand{\arraystretch}{1.12}
\setlength{\tabcolsep}{4pt}

\definecolor{DeepBlue}{RGB}{77,88,234}
\definecolor{LightBlue}{RGB}{235,241,250}
\renewcommand{\chg}[1]{\textcolor{gray}{\scriptsize\,({#1})}}
\newcommand{\best}[1]{\textcolor{DeepBlue}{\textbf{#1}}}

\begin{tabular}{c|c|ccc}
\toprule
\textbf{Model} & \textbf{Attempt} & \textbf{Func.} & \textbf{Non-func.} & \textbf{Reuse (\%)} \\
\hline
\multirow{4}{*}{\shortstack{GPT-5\\-2025-08-07}}
& A1 & 58.20 & 65.33 & -- \\
& A2 & 66.91\chg{+8.71} & 65.28\chg{-0.05} & 44.74 \\
& A3 & 71.62\chg{+13.42} & 66.86\chg{+1.53} & 60.53 \\
& \cellcolor{LightBlue}\textcolor{DeepBlue}{\textbf{A4}}
& \cellcolor{LightBlue}\best{72.32}\chg{+14.12}
& \cellcolor{LightBlue}\best{65.58}\chg{+0.25}
& \cellcolor{LightBlue}\best{63.16} \\
\hline
\multirow{4}{*}{\shortstack{DeepSeek\\-V3-0324}}
& A1 & 23.99 & 55.24 & -- \\
& A2 & 34.37\chg{+10.38} & 54.24\chg{-1.00} & 10.53 \\
& A3 & 35.69\chg{+11.70} & 56.32\chg{+1.08} & 18.42 \\
& \cellcolor{LightBlue}\textcolor{DeepBlue}{\textbf{A4}}
& \cellcolor{LightBlue}\best{36.25}\chg{+12.26}
& \cellcolor{LightBlue}\best{57.32}\chg{+2.08}
& \cellcolor{LightBlue}\best{31.58} \\
\hline
\multirow{4}{*}{\shortstack{Claude-Sonnet\\-4.5-20250929}}
& A1 & 67.01 & 64.81 & -- \\
& A2 & 81.65\chg{+14.64} & 69.49\chg{+4.68} & 42.11 \\
& A3 & 82.88\chg{+15.87} & 68.35\chg{+3.54} & 57.89 \\
& \cellcolor{LightBlue}\textcolor{DeepBlue}{\textbf{A4}}
& \cellcolor{LightBlue}\best{85.73}\chg{+18.72}
& \cellcolor{LightBlue}\best{69.76}\chg{+4.95}
& \cellcolor{LightBlue}\best{63.16} \\
\hline
\multirow{4}{*}{\shortstack{Gemini-3\\-Pro-Preview}}
& A1 & 69.16 & 67.03 & -- \\
& A2 & 90.50\chg{+21.34} & 69.08\chg{+2.05} & 57.89 \\
& A3 & 90.70\chg{+21.54} & 69.46\chg{+2.43} & 73.68 \\
& \cellcolor{LightBlue}\textcolor{DeepBlue}{\textbf{A4}}
& \cellcolor{LightBlue}\best{92.10}\chg{+22.94}
& \cellcolor{LightBlue}\best{65.71}\chg{-1.32}
& \cellcolor{LightBlue}\best{81.58} \\
\bottomrule
\end{tabular}
\end{table}

% \lh{comments about the metric.}

To investigate the impact of knowledge evolution across attempts, we conduct studies and report the results in Table~\ref{tab:rq2_knowledge_evolution}.
Overall, the functional score improves monotonically from A1 to A4 for all four backbone models, with absolute gains of 14.12 points for GPT-5, 12.26 for DeepSeek-V3, 18.72 for Claude-Sonnet-4.5, and 22.94 for Gemini-3-Pro-Preview. 
At the same time, repository reuse also increases steadily across attempts, reaching 63.16\%, 31.58\%, 63.16\%, and 81.58\%, respectively, in A4. 
The joint increase in functional performance and repository reuse suggests that later attempts increasingly build on accumulated task-specific state.

\textit{Attempt-wise improvement.}
The largest gains generally appear early. From A1 to A2 alone, the functional score increases by 8.71 points for GPT-5, 10.38 for DeepSeek-V3, 14.64 for Claude-Sonnet-4.5, and 21.34 for Gemini-3-Pro-Preview. 
Later attempts continue to improve performance, although with smaller increments, suggesting diminishing returns rather than random fluctuation. 
This pattern suggests that knowledge evolution across attempts is effective for both weaker and stronger backbones: DeepSeek-V3 improves substantially from a much lower starting point, while Claude-Sonnet-4.5 and Gemini-3-Pro-Preview still achieve large gains despite already strong A1 results.

\textit{Reuse and quality stability.}
The repository reuse rate increases monotonically for all four models, suggesting that later attempts increasingly preserve and exploit previously validated repository content. 
This trend is especially pronounced for Claude-Sonnet-4.5 and Gemini-3-Pro-Preview, whose reuse rates rise to 63.16\% and 81.58\%, respectively, while DeepSeek-V3 shows a lower but still steady increase from 10.53\% to 31.58\%. 
Meanwhile, non-functional quality remains largely stable across attempts. Compared with A1, the final non-functional score is slightly higher for GPT-5, DeepSeek-V3, and Claude-Sonnet-4.5, and only slightly lower for Gemini-3-Pro-Preview. More specifically, GPT-5 changes from 65.33 to 65.58, DeepSeek-V3 from 55.24 to 57.32, Claude-Sonnet-4.5 from 64.81 to 69.76, and Gemini-3-Pro-Preview from 67.03 to 65.71 after improving in A2 and A3.
Overall, the substantial functional gains of \approach do not coincide with a general deterioration in non-functional quality.

\begin{tcolorbox}[
  enhanced,
  colback=grey,                % 背景色
  colframe=teal!60!black,       % 边框颜色
  boxrule=0.6pt,                % 边框线宽
  arc=10pt,                     % 圆角大小
  left=4mm,right=4mm,           % 内边距
  top=2mm,bottom=2mm,
  drop shadow={black!40!white}, % 右下角阴影
]
\textbf{\textit{Answer to RQ2:} }
\textit{Knowledge evolution across attempts consistently improves functional performance across all backbone models, while steadily increasing repository reuse and keeping non-functional quality broadly stable. These results suggest that later attempts increasingly build on accumulated task-specific state rather than acting as independent retries.}

\end{tcolorbox}

% \begin{table}[t]
% \centering
% \caption{Impact of iterative knowledge evolution on RAL-Bench. 
% We report the functional score (Func.), non-functional score (Non-func.), cost, and repository reuse rate across knowledge evolution rounds.}
% \label{tab:rq2_knowledge_evolution}
% \small
% \renewcommand{\arraystretch}{1.12}
% \setlength{\tabcolsep}{4pt}
% \begin{tabular}{c|c|cccc}
% \toprule
% \textbf{Model} & \textbf{Round} & \textbf{Func.} & \textbf{Non-func.} & \textbf{Cost (\$)} & \textbf{Reuse (\%)} \\
% % \midrule
% \hline
% \multirow{4}{*}{\shortstack{GPT-5\\2025-08-07}}
% & A1 & 58.20 & 65.33 & 13.24 & -- \\
% & A2 & 66.91 & 65.28 & 7.58 & 44.74 \\
% & A3 & 71.62 & 66.86 & 7.42 & 60.53 \\
% & A4 & 72.32 & 65.58 & 7.31 & 63.16 \\
% % \midrule
% \hline
% \multirow{4}{*}{\shortstack{DeepSeek\\V3-0324}}
% & A1 & 23.99 & 55.24 & 6.43 & -- \\
% & A2 & 34.37 & 54.24 & 4.36 & 10.53 \\
% & A3 & 35.69 & 56.32 & 4.02 & 18.42 \\
% & A4 & 36.25 & 57.32 & 3.73 & 31.58 \\
% % \midrule
% \hline
% \multirow{4}{*}{\shortstack{Claude-Sonnet\\4.5-20250929}}
% & A1 & 67.01 & 64.81 & 112.45 & -- \\
% & A2 & 81.65 & 69.49 & 87.43 & 42.11 \\
% & A3 & 82.88 & 68.35 & 77.39 & 57.89 \\
% & A4 & 85.73 & 69.76 & 52.14 & 63.16 \\
% % \midrule
% \hline
% \multirow{4}{*}{\shortstack{Gemini-3\\Pro-Preview}}
% & A1 & 69.16 & 67.03 & 54.23 & -- \\
% & A2 & 90.50 & 69.08 & 33.71 & 57.89 \\
% & A3 & 90.70 & 69.46 & 30.65 & 73.68 \\
% & A4 & 92.10 & 65.71 & 25.72 & 81.58 \\
% \bottomrule
% \end{tabular}
% \end{table}

\subsection{RQ3: Ablation Study}

\begin{table}[t]
\centering
\caption{Ablation study of \approach on RAL-Bench across backbone models. Ablation is conducted at Attempt 2 (A2), i.e., the first attempt that can leverage accumulated cross-attempt state.
All numbers are percentages. Gray numbers in parentheses show the absolute change from \approach under the same backbone model.}
\label{tab:ablation_ralbench}
\small
\renewcommand{\arraystretch}{1.12}
\setlength{\tabcolsep}{4pt}

\definecolor{DeepBlue}{RGB}{77,88,234}
\definecolor{LightBlue}{RGB}{235,241,250}
\renewcommand{\chg}[1]{\textcolor{gray}{\scriptsize\,({#1})}}
\newcommand{\best}[1]{\textcolor{DeepBlue}{\textbf{#1}}}

\begin{tabular}{c|l|cc}
\toprule
\textbf{Model} & \textbf{Variant} & \textbf{Func.} & \textbf{Non-func.} \\
\hline

\multirow{4}{*}{\centering\arraybackslash\shortstack{GPT-5\\-2025-08-07}}
& \cellcolor{LightBlue}\textcolor{DeepBlue}{\textbf{\approach}}
& \cellcolor{LightBlue}\best{66.91}
& \cellcolor{LightBlue}\best{65.28} \\
& w/o Success Knowledge
& 63.72\chg{-3.19}
& 64.32\chg{-0.96} \\
& w/o Failure Knowledge
& 61.84\chg{-5.07}
& 61.19\chg{-4.09} \\
& w/o Historical-Best Repo.
& 65.47\chg{-1.44}
& 66.93\chg{+1.65} \\
\hline

\multirow{4}{*}{\centering\arraybackslash\shortstack{DeepSeek\\-V3-0324}}
& \cellcolor{LightBlue}\textcolor{DeepBlue}{\textbf{\approach}}
& \cellcolor{LightBlue}\best{34.37}
& \cellcolor{LightBlue}\best{44.24} \\
& w/o Success Knowledge
& 11.50\chg{-22.87}
& 41.80\chg{-2.44} \\
& w/o Failure Knowledge
& 18.26\chg{-16.11}
& 46.79\chg{+2.55} \\
& w/o Historical-Best Repo.
& 15.93\chg{-18.44}
& 50.47\chg{+6.23} \\
\hline

\multirow{4}{*}{\centering\arraybackslash\shortstack{Claude-Sonnet\\-4.5-20250929}}
& \cellcolor{LightBlue}\textcolor{DeepBlue}{\textbf{\approach}}
& \cellcolor{LightBlue}\best{81.65}
& \cellcolor{LightBlue}\best{69.49} \\
& w/o Success Knowledge
& 45.12\chg{-36.53}
& 65.43\chg{-4.06} \\
& w/o Failure Knowledge
& 77.05\chg{-4.60}
& 66.39\chg{-3.10} \\
& w/o Historical-Best Repo.
& 72.77\chg{-8.88}
& 64.70\chg{-4.79} \\
\hline

\multirow{4}{*}{\centering\arraybackslash\shortstack{Gemini-3\\-Pro-Preview}}
& \cellcolor{LightBlue}\textcolor{DeepBlue}{\textbf{\approach}}
& \cellcolor{LightBlue}\best{90.51}
& \cellcolor{LightBlue}\best{69.09} \\
& w/o Success Knowledge
& 86.61\chg{-3.90}
& 68.31\chg{-0.78} \\
& w/o Failure Knowledge
& 89.82\chg{-0.69}
& 66.56\chg{-2.53} \\
& w/o Historical-Best Repo.
& 78.93\chg{-11.58}
& 67.31\chg{-1.78} \\
\bottomrule
\end{tabular}
\end{table}

To analyze the individual contributions of the three persistent-state components in \approach, we conduct ablation studies on RAL-Bench in Table~\ref{tab:ablation_ralbench}. 
"w/o Success Knowledge" removes the reuse of success knowledge, so subsequent attempts can no longer leverage signals from previously strong repositories. 
"w/o Failure Knowledge" disables the use of failure knowledge, so subsequent attempts no longer receive explicit signals from prior unsuccessful outcomes. 
"w/o Historical-Best Repo." removes the preservation of the historical-best repository, so subsequent attempts proceed without an explicit safeguard against regression.

\textit{Component roles.}
The ablation results reveal clear model-dependent differences, while consistently showing that all three persistent task-specific state components contribute to the effectiveness of \approach. Success Knowledge is the most critical component for Claude-Sonnet-4.5 and DeepSeek-V3, where its removal causes the largest functional drops of 36.53 and 22.87 points, respectively. This result suggests that these two backbones benefit most from reusing success knowledge from previously strong repositories. In contrast, GPT-5 is most sensitive to the removal of Failure Knowledge, with the largest declines in both functional and non-functional scores, indicating that explicit failure signals are particularly important for avoiding previously ineffective directions. For Gemini-3-Pro-Preview, removing Historical-Best Repository causes the largest functional drop, suggesting that preserving the historical-best repository is especially important for mitigating regression across attempts.

\textit{Metric comparison.}
The non-functional results show a more mixed pattern than the functional results. In several cases, removing a component slightly improves the non-functional score while reducing the functional score; for example, removing Historical-Best Repository increases the non-functional score for GPT-5 and DeepSeek-V3. This result suggests that functional correctness and non-functional quality are not always aligned. Nevertheless, the consistent functional degradation across all ablations confirms that each component remains necessary for stable repository-level generation across attempts.

\begin{tcolorbox}[
  enhanced,
  colback=grey,
  colframe=teal!60!black,
  boxrule=0.6pt,
  arc=10pt,
  left=4mm,right=4mm,
  top=2mm,bottom=2mm,
  drop shadow={black!40!white},
]
\textbf{\textit{Answer to RQ3:} }
\textit{All three persistent task-specific state components contribute to \approach. Their effects are complementary but model-dependent, and removing any one consistently reduces functional performance. This result shows that effective repository-level generation across attempts requires success knowledge, failure knowledge, and preservation of the historical-best repository.}
\end{tcolorbox}

\subsection{RQ4: Costs and Error Analysis}

\begin{figure}[t]
    \centering
    \includegraphics[width=0.8\linewidth]{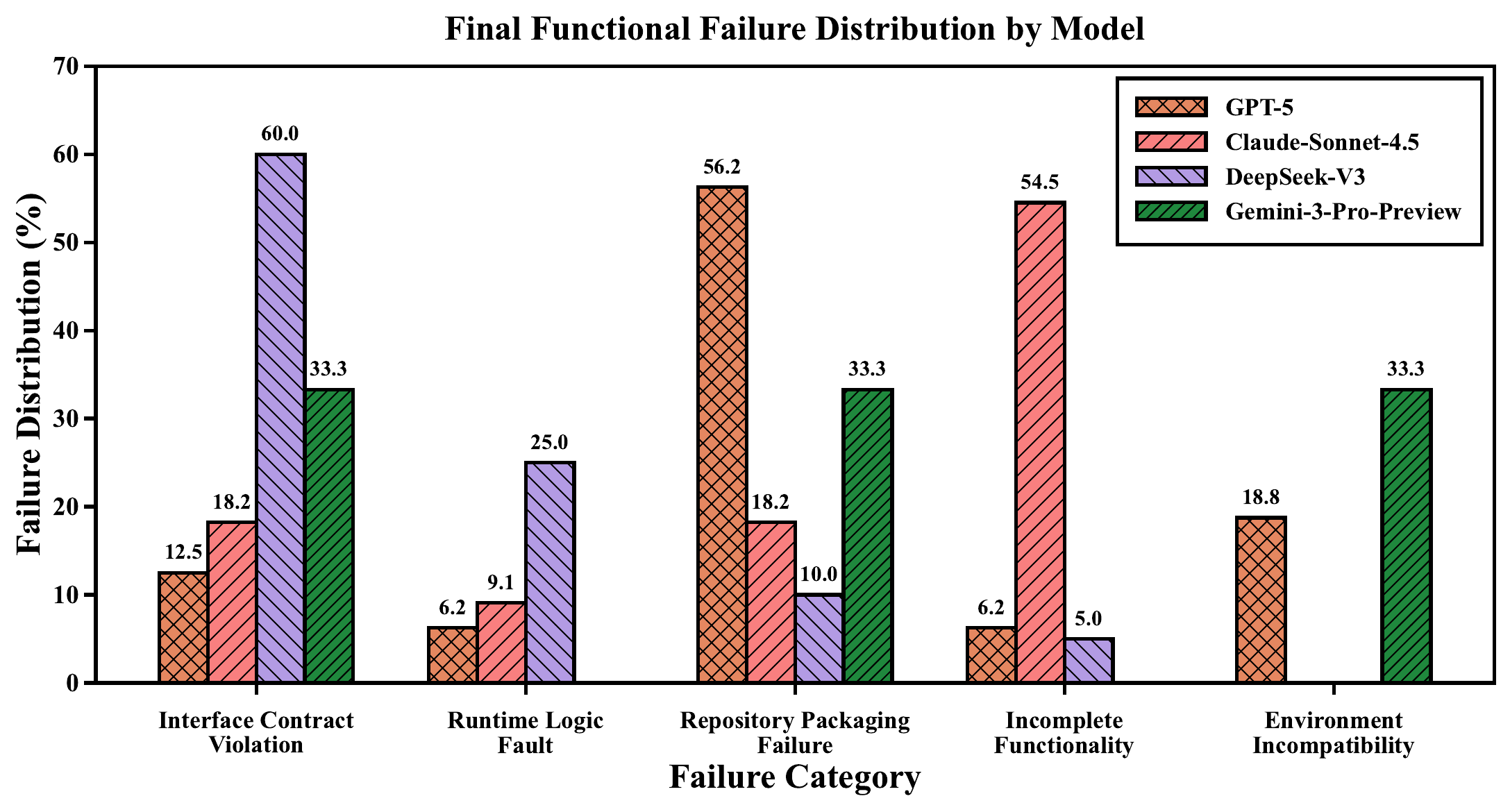}
    \vspace{-5px}
    \caption{Distribution of residual functional failures after knowledge evolution across attempts.
    }
    \label{fig:rq4_failure_distribution}
    % \vspace{-5px}
\end{figure}

\begin{table}[t]
\centering
\caption{Cost trajectory of knowledge evolution across attempts on RAL-Bench. We report the cost of each attempt (in \$) and the overall reduction from Attempt~1 to Attempt~4, with gray numbers indicating the absolute decrease relative to Attempt~1.}
\label{tab:rq4_cost_trajectory}
\small
\renewcommand{\arraystretch}{1.02}
\setlength{\tabcolsep}{2.8pt}

\begin{tabular}{>{\centering\arraybackslash}p{1.6cm}
                >{\centering\arraybackslash}p{1.03cm}
                >{\centering\arraybackslash}p{1.28cm}
                >{\centering\arraybackslash}p{1.28cm}
                >{\centering\arraybackslash}p{1.28cm}
                >{\centering\arraybackslash}p{0.9cm}}
\toprule
\textbf{Model} &
\textbf{Attempt~1} &
\textbf{Attempt~2} &
\textbf{Attempt~3} &
\textbf{Attempt~4} &
\textbf{\textcolor{DeepRed}{Reduc.}} \\
\hline
\makecell[c]{\textbf{GPT-5}\\\textbf{-2025-08-07}}
& 13.24
& 7.58\,\costabs{-5.7}
& 7.42\,\costabs{-5.8}
& 7.31\,\costabs{-5.9}
& \cellcolor{LightRed}\redimp{44.79\%} \\
\hline
\makecell[c]{\textbf{DeepSeek}\\\textbf{-V3-0324}}
& 6.43
& 4.36\,\costabs{-2.1}
& 4.02\,\costabs{-2.4}
& 3.73\,\costabs{-2.7}
& \cellcolor{LightRed}\redimp{41.99\%} \\
\hline
\makecell[c]{\textbf{Claude}\\\textbf{-Sonnet-4.5}}
& 112.45
& 87.43\,\costabs{-25.0}
& 77.39\,\costabs{-35.1}
& 52.14\,\costabs{-60.3}
& \cellcolor{LightRed}\redimp{53.63\%} \\
\hline
\makecell[c]{\textbf{Gemini-3}\\\textbf{-Pro-Preview}}
& 54.23
& 33.71\,\costabs{-20.5}
& 30.65\,\costabs{-23.6}
& 25.72\,\costabs{-28.5}
& \cellcolor{LightRed}\redimp{52.57\%} \\
\bottomrule
\end{tabular}
\end{table}

For this RQ, we further investigate the usage of \approach:

\textit{Cost analysis.}
We perform a cost analysis for \approach from Attempt~1 to Attempt~4 on RAL-Bench to examine whether the gains of cross-attempt optimization are obtained at the expense of increasing retry cost. We estimate monetary cost using model-specific API pricing and sum all LLM calls made within each attempt. Table~\ref{tab:rq4_cost_trajectory} reports the cost of each attempt together with the relative reduction from Attempt~1 to Attempt~4 under four backbone models. 
As shown in the table, while the performance of \approach improves across attempts, the cost consistently decreases for all four models, dropping from 13.24 to 7.31 for GPT-5, from 6.43 to 3.73 for DeepSeek-V3, from 112.45 to 52.14 for Claude-Sonnet-4.5, and from 54.23 to 25.72 for Gemini-3-Pro-Preview, corresponding to relative reductions of 44.79\%, 41.99\%, 53.63\%, and 52.57\%, respectively. 
These results indicate that the gains of \approach are not obtained by simply spending more on repeated retries. 
Instead, as task-specific knowledge accumulates across attempts, later attempts can increasingly reuse validated repository artifacts and narrow the repair search space, thereby improving effectiveness while reducing redundant generation cost.

\textit{Error analysis.}
We analyze the residual errors of \approach on RAL-Bench across four backbone models. We manually inspect the final failed repositories returned after Attempt~4 and assign each case one primary category based on the dominant issue reflected in the runtime logs and test outcomes. We group the residual failures into five categories: \emph{Interface Contract Violation}, \emph{Runtime Logic Fault}, \emph{Repository Packaging Failure}, \emph{Incomplete Functionality}, and \emph{Environment Incompatibility}. Fig.~\ref{fig:rq4_failure_distribution} shows that the remaining failures are clearly model-dependent and are concentrated in deeper functional and repository-integration bottlenecks rather than a single early-stage issue. GPT-5 is dominated by \emph{Repository Packaging Failure}, suggesting that its main residual weakness lies in repository organization and executable packaging. Claude-Sonnet-4.5 is dominated by \emph{Incomplete Functionality}, indicating that many failed cases can build and run but still do not fully implement the required behaviors. DeepSeek-V3 is primarily affected by \emph{Interface Contract Violation}, with an additional share of \emph{Runtime Logic Fault}, suggesting difficulty in maintaining cross-file consistency and correct execution behavior. Gemini-3-Pro-Preview exhibits a more balanced distribution across \emph{Interface Contract Violation}, \emph{Repository Packaging Failure}, and \emph{Environment Incompatibility}, indicating broader repository-level integration challenges. Overall, the remaining bottlenecks lie in packaging robustness, behavior completion, contract preservation, and end-to-end integration. Further improvements should therefore strengthen cross-file reasoning, behavior-level validation, and backbone-adaptive repair.

\begin{tcolorbox}[
  enhanced,
  colback=grey,                % 背景色
  colframe=teal!60!black,       % 边框颜色
  boxrule=0.6pt,                % 边框线宽
  arc=10pt,                     % 圆角大小
  left=4mm,right=4mm,           % 内边距
  top=2mm,bottom=2mm,
  drop shadow={black!40!white}, % 右下角阴影
]
\textbf{\textit{Answer to RQ4:} }
\textit{\approach remains practical under cross-attempt knowledge evolution, achieving consistent cost reduction across backbone models. The remaining failures mainly lie in deeper and more model-specific bottlenecks, particularly behavior completion, contract preservation, and repository-level integration.}

\end{tcolorbox}

\subsection{Threats to Validity}

\textbf{Threats in the evaluation protocol.}
A primary threat concerns the evaluation protocol. 
We adopt a unified prompting strategy and a fixed attempt budget to ensure comparability across methods and backbone models. However, alternative prompts, different attempt budgets, or adaptive stopping criteria may change the absolute scores. This limitation is particularly relevant to repeated repository generation, because some tasks may saturate early whereas others may still benefit from additional attempts. Even so, the same protocol is applied to all compared methods, and the improvements of \approach remain stable under this controlled setting. We leave the exploration of adaptive prompting and attempt scheduling to future work.

\textbf{Threats in generalizability.}
Another potential threat relates to the generalizability of our evaluation. 
Although we evaluate \approach on two representative repository-level benchmarks and four frontier backbone models, the current study still covers only a subset of task types and a single programming language. To mitigate this threat, we report both functional and non-functional results and examine repeated generation across multiple attempts under a unified evaluation protocol. The consistently positive trends across backbone models suggest that the conclusions are not tied to a single benchmark or a single model family. Nevertheless, we do not claim that the current evaluation fully covers all repository-level generation settings. In future work, we plan to further validate the generalizability of \approach on broader benchmark collections, more diverse software settings, and additional backbone models.

\textbf{Threats in benchmark and model contamination.}
A third potential threat comes from benchmark and model contamination. Because modern backbone models are trained on large-scale public code and web corpora, it is difficult to completely rule out overlap between the pre-training data and benchmark repositories, dependencies, or common implementation patterns. 
However, this threat affects all compared methods under the same backbone model and benchmark setting. Therefore, although contamination may influence the absolute performance level, it does not compromise the fairness of our comparative analysis or the relative gains of \approach, which remain stable under the same evaluation settings.

\section{Related Work}

Recent advances in LLMs have substantially improved code generation across a wide range of settings \citep{jiang2026survey, liu2023your, li2023starcoder, lozhkov2024starcoder, nam2024using, joel2024survey}. Early studies mainly focused on function-level generation \citep{du2024evaluating, chen2021evaluating, austin2021program}, where models generate individual functions from natural language descriptions. Subsequent work extended this paradigm to more challenging settings, such as competitive programming \citep{hendrycks2021measuring, li2022competition, islam2024mapcoder}. More recently, research has shifted toward repository-level code generation, including repository completion \citep{zhang2023repocoder, li2024deveval}, feature implementation \citep{deng2025nocode, li2025fea}, and generation of complete repositories from scratch \citep{pan2026ral, luo2025rpg}, where models must coordinate multiple files, interfaces, and dependencies within a codebase. Compared with earlier settings, repository-level code generation more closely reflects real-world software development and often involves repeated development cycles, such as iterative submissions, revisions, and continued refinement. This characteristic has motivated growing interest in methods that exploit information from prior generations or iterative feedback, rather than treating each generation process as fully isolated \citep{pan2025codecor, bi2024iterative}.

A major line of research improves code generation through iterative and self-improving strategies \citep{defferrard2024towards}. Rather than relying on single-pass generation, these approaches progressively refine outputs through reflection, search, evaluation feedback, or interaction with development environments. For example, Self-Reflection \citep{madaan2023self} generates natural-language reflections on previous outputs to guide subsequent iterations, whereas SE-Agent \citep{lin2025se} improves performance by revising and recombining prior reasoning trajectories. AlphaEvolve \citep{novikov2025alphaevolve} iteratively mutates candidate programs and selects stronger ones based on evaluation signals, while CSE \citep{hu2026controlled} performs collaborative self-evolution over multiple candidate solutions. Live-SWE-Agent \citep{xia2025live} further extends such iterative optimization to repository-level software engineering by enabling agents to interact with tools and execution environments for repository modification and repair. These approaches demonstrate the effectiveness of iterative improvement for complex code generation tasks, but they still primarily optimize the current generation process, candidate set, or reasoning trajectory.

However, repeated repository-level code generation requires more than within-attempt optimization. In real-world software development, the same task may be revisited through multiple cycles of generation, revision, and resubmission, making it insufficient to treat each attempt as an isolated optimization episode \citep{zhang2026evocodebench, zhang2024codeagent}. The central difficulty lies not only in improving the current repository, but also in preserving and reusing task-specific experience across attempts. Existing approaches generally optimize the current generation process, candidate set, or reasoning trajectory, but do not explicitly preserve reusable knowledge from prior successes and failures for the same task. They also lack explicit mechanisms for directly reusing a previously strong repository, avoiding known failure patterns, or preventing newly generated solutions from replacing a stronger historical best. As a result, they may repeatedly incur similar exploration costs and even regress to inferior repositories. Our work instead focuses on persistent cross-attempt experience optimization. Specifically, it explicitly maintains reusable success and failure knowledge across repeated attempts, together with the historical-best repository. This distinction separates our work from prior iterative refinement, search-based, and agentic approaches that mainly optimize within a single attempt.

\section{Conclusion}

In this paper, we propose the \approach framework, which explicitly models repository-level code generation as a cross-attempt optimization process with persistent task-specific state. 
It preserves and reuses three key forms of accumulated state across attempts, namely success knowledge, failure knowledge, and the historical-best repository, thereby turning repeated generation from isolated retries into a knowledge-driven optimization process. 
Extensive experiments on representative benchmarks and multiple frontier LLMs demonstrate the superior effectiveness, stability, and cost--effectiveness of \approach for repository-level executable code generation. 
Our work represents a promising step towards more persistent, adaptive, and efficient repository-level code generation beyond single-attempt refinement. 
In future work, we plan to further extend \approach to broader benchmark collections, more diverse software stacks, and stronger backbone models. We will also explore applications of the \approach framework to other domains beyond code generation. 
% All related scripts and data are available at \href{https://anonymous.4open.science/r/LiveCoder}
%      {\textbf{\textcolor{oai-orange}{\nolinkurl{https://anonymous.4open.science/r/LiveCoder}}}}.
% All related scripts and data are available at \href{https://anonymous.4open.science/r/LiveCoder}
% {\textbf{\textcolor{oai-orange}{\nolinkurl{https://anonymous.4open.science/r/LiveCoder}}}} and are also archived in \href{https://zenodo.org/records/19243137}
% {\textbf{\textcolor{oai-orange}{an anonymous Zenodo repository}}}.

% \section{Data Availability}

% To ensure the reproducibility of our results, all related scripts and data are available at \href{https://anonymous.4open.science/r/LiveCoder}
% {\textbf{\textcolor{oai-orange}{\nolinkurl{https://anonymous.4open.science/r/LiveCoder}}}} and are also archived in \href{https://zenodo.org/records/19243137}
% {\textbf{\textcolor{oai-orange}{an anonymous Zenodo repository}}}.

% To ensure the reproducibility of our results, all related scripts and data are available at \href{https://github.com/Wwstarry/LiveCoder}{\textbf{\textcolor{oai-orange}{the repository}}}.

\bibliographystyle{ACM-Reference-Format}
\bibliography{references}

\end{document}